\documentclass[preprint,showpacs,prb,superscriptaddress]{revtex4}
\usepackage{amssymb,amsmath}
\usepackage[dvips]{graphicx}

\usepackage{dcolumn,epsfig}

\newcommand{\beq}{\begin{equation}}
\newcommand{\eeq}{\end{equation}}
\newcommand{\bes}{\begin{subequations}}
\newcommand{\ees}{\end{subequations}}
\newcommand{\bea}{\begin{eqnarray}}
\newcommand{\eea}{\end{eqnarray}}
\newcommand{\ba}{\begin{array}}
\newcommand{\ea}{\end{array}}
\newcommand{\beqn}{\begin{eqnarray*}}
\newcommand{\eeqn}{\end{eqnarray*}}
        
\newcommand{\p}{\partial}
\newcommand{\f}[2]{\frac{#1}{#2}}
\newcommand{\g}{\gamma}
\newcommand{\al}{\alpha}
\newcommand{\n}{\eta}
\newcommand{\tc}{\tilde{c}}
\newcommand{\tn}{\tilde{\eta}}
\newcommand{\Se}{\Sigma}
\newcommand{\om}{\omega}
\newcommand{\G}{\Gamma}
\newcommand{\la}{\langle}
\newcommand{\ra}{\rangle}
\newcommand{\mT}{\mathcal{T}} 
\newcommand{\mF}{\mathcal{F}}

\def\nn{\nonumber}

\newlength{\sizeonefig}
\newlength{\sizetwofig}
\begin{document}

\title{Electron transport in a one dimensional conductor with
  inelastic scattering by self-consistent reservoirs}

\author{Dibyendu Roy} 
\email{dibyendu@rri.res.in}
\author{Abhishek Dhar} 
\email{dabhi@rri.res.in}
\affiliation{Raman Research Institute, Bangalore 560080}

\begin{abstract}
We present an extension of the work of D'Amato and Pastawski on
electron transport in a one-dimensional conductor modeled by 
the tight binding lattice Hamiltonian and in which inelastic 
scattering is incorporated by connecting each site of the lattice to
one-dimensional leads. This model incorporates B\"uttiker's original
idea of dephasing probes.  
Here we consider finite temperatures and study
both electrical and heat transport across a chain with applied
chemical potential and temperature gradients. Our approach involves
quantum Langevin equations and nonequilibrium Green's functions.
In the linear response limit we are able to solve the model exactly
and obtain expressions for various transport coefficients. Standard 
linear response relations are shown to be 
valid. We also explicitly compute the heat dissipation and show that for wires of length $N >>
\ell$, where $\ell$ is a coherence length scale, dissipation takes
place uniformly along the wire. For $N << \ell$, when transport is ballistic,
dissipation is mostly at the contacts. In the intermediate range
between Ohmic and ballistic transport we find that the chemical
potential profile is linear in the bulk with sharp jumps at the
boundaries. These are explained using a simple model where the left
and right moving electrons behave as persistent random walkers. 
\end{abstract}

\vspace{0.5cm}
\date{\today}

\maketitle

\section{Introduction}
Inelastic scattering provides a mechanism for dissipation and
decoherence in quantum systems. These effects are important in
considering transport properties of mesoscopic systems. Experimental
examples are numerous and include studies of transport in systems such
as  single walled carbon
nanotubes \cite{Tans97}, atomic chains \cite{Ohnishi98, Yanson98},
semiconducting heterostructures \cite{dePicciotto01} and polymer
nanofibers \cite{Zimbov05}. 
In the absence of inelastic scattering, 
transport is either ballistic and we see effects such as conductance
quantization \cite{VanWees88, Wharam88}, or, with elastic scatterers
we see effects of coherent 
scattering such as Anderson localization \cite{navarro05}. In either
case transport is 
non-Ohmic even when we consider very long wires. Introducing inelastic
scattering necessarily leads to decoherence and both of the above
effects (ballistic transport, localization) are reduced. One expects
that in the limit of long wires one should get Ohmic transport
\cite{Datta05}. Recent experiments on 
atomic chains \cite{Agrait02} and Fullerene bridges \cite{Park00,
  Pasupathy05} have studied the effects of inelastic scattering and
the associated local heating on quantum transport.   

The physical sources for inelastic scattering are well known and occur
basically due to the interaction of the conducting electrons with
other degrees of freedom in the system. For example these could arise
due to electron-phonon interactions or interactions between conducting
and non-conducting electrons \cite{Ziman}. However the microscopic modeling of
inelastic scattering in the context of transport is
nontrivial. 
One of the first phenomenological models for dissipation was due to
B\"uttiker \cite{Buttiker85, Buttiker86}. 
In B\"uttiker's model one connects a point inside the wire to a
reservoir of electrons 
maintained at a chemical potential $\mu$ whose value is set by the
condition that there is no average current flow into this side
reservoir. This is equivalent to connecting a voltage probe at some
point on the wire and a nice experimental realization of this
situation can be seen in [\onlinecite{dePicciotto01}]. In B\"uttiker's model
an electron flowing into the reservoir can emerge with a different
phase and energy and thus one can have both decoherence and dissipation. 

A more detailed microscopic calculation using B\"uttiker's idea of
incorporating inelastic scattering was performed by D'Amato and
Pastawski \cite{Pastawski90}. In their study they  
considered transmission across a wire modeled by the tight-binding Hamiltonian with a 
nearest-neighbour hopping parameter $V$. Each site on 
the wire is connected to electron baths which are themselves modeled by
tight-binding Hamiltonians with hopping parameter $\eta$. The wire is
attached at  the two ends to ideal leads  
 with the same hopping parameter as the wire. These two leads are
 connected to reservoirs kept at fixed chemical 
potentials  $\mu_L$ and $\mu_R$ for the left and right leads respectively. The side 
leads are attached to reservoirs whose chemical potentials are fixed self-consistently
by imposing the condition of zero current. Using this model 
D'Amato and Pastawski analytically solved the case where the
self-energy correction due to the side leads is pure imaginary and has
the form $i \eta$ and $\eta$ is small.  
They were able to demonstrate the transition from coherent to Ohmic transport.
An inelastic  length scale $\ell=aV/\eta$, with $a$ as a lattice
parameter, was  
introduced such that  for wire length $L<<\ell$ transport was coherent while for $L >> \ell$ transport 
was Ohmic. A number of other papers \cite{Maschke91, Datta89, Datta91}
 have also shown that other 
models of inelastic scattering, for example due to electron-phonon (using side reservoirs as ensemble of harmonic oscillators to describe the heat bath) or 
electron-electron interactions, can be related to the B\"uttiker
mechanism. Some recent papers have looked at electron-phonon 
interactions using the Keldysh nonequilibrium Green's function formalism 
combined with density-functional methods \cite{Frederiksen04},
tight-binding molecular-dynamics \cite{Yamamoto05} and the self-consistent
Born approximation \cite{Asai04}. An alternative mechanism for
introducing inelastic scattering, through introduction of an imaginary
potential in the Hamiltonian, has  also been studied \cite{efetov95,mccann96,brouwer97,arun02}.  

In the present paper we present an extension of the work of D'Amato
and Pastawski. We study  
the case of transport of both heat and electron in the presence of inelastic scatterers in 
the form of self-consistent leads. 
The wire
is subjected to both chemical potential and temperature gradients and
we evaluate steady state values of both the particle and heat 
current operators. In the limit of a long wire when one is in the
Ohmic regime we are able to obtain explicit expressions for all the
linear response coefficients. It is verified that various linear
response results such as Onsager reciprocity and the Weidemann-Franz
law are valid. In the intermediate regime between ballistic and Ohmic
transport we propose a simple model of right moving and left moving
persistent random walkers  which can explain much of the observed behaviour.
We also perform an explicit calculation of the
heat loss along the wire. This is a second order effect in the
gradients and we show that there is uniform heat dissipation along the
length of the wire whose value is  precisely the Joule heat loss.
For short wires we show that heat dissipation takes place
primarily at the contacts.  
While heat dissipation by B\"uttiker probes has been discussed
in [\onlinecite{Buttiker86,DattaLake92}],  
we believe that this is the first explicitly microscopic
calculation of dissipation in a quantum wire that clearly
demonstrates  Joule heat loss in the Ohmic regime and dissipation into
the reservoirs in the ballistic regime. 

The formalism used in this paper is the quantum Langevin equations
approach. In two  recent papers \cite{AbhishekShastry03,
  AbhishekSen06} it was shown how this approach can be used 
to derive both the Landauer results and more generally the
nonequilibrium Green's function (NEGF) results on transport. 
Here we show how this method also works for
the mulitple reservoir case and quickly leads to NEGF-like expressions
for currents for both particle and heat. These equations are the
starting point of our analysis. Thus apart from extending the results
of [\onlinecite{Pastawski90}] we also use a different and more general
approach. Unlike [\onlinecite{Pastawski90}] we also consider large
values of the inelasticity parameter.

The paper is organized as follows. In sec.~(\ref{sec:model}) we define the
model and describe how the quantum Langevin approach can be used to
get formal expressions for electron and heat currents in the steady
state. In sec.~(\ref{sec:selfcons}) we write the self-consistent
equations and discuss the linear response regime. In
Sec.~(\ref{sec:solu}) we solve the self-consistent equations for a long
wire which is kept in a specified temperature gradient and evaluate
the electical and heat current along the wire and also the heat loss
into the side reservoirs. The transition from the ballistic to the
Ohmic regime is briefly discussed in sec.~(\ref{sec:finite}). Finally
we conclude with a discussion in sec.~(\ref{sec:disc}).   

\section{Model and general results}
\label{sec:model}

\begin{figure}[t]
\begin{center}
\includegraphics[width=16.0cm]{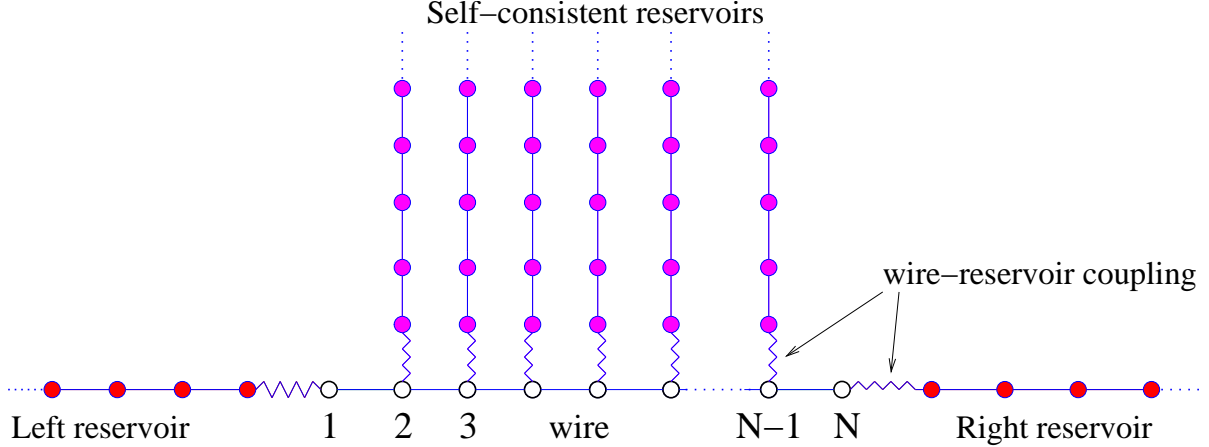}
\end{center}
\caption{ A schematic description of the model. }
\label{cartoon}
\end{figure}

We consider a one-dimensional wire modeled by the
tight-binding lattice Hamiltonian. The wire has  $N$ sites each of
which is coupled to an infinite reservoir which is itself
modeled by a one-dimensional tight-binding system [see Fig.~(\ref{cartoon})]. 
The Hamiltonian of the system consisting of the wire and all
the reservoirs  is given by
\begin{eqnarray}
\mathcal{H} &=& \mathcal{H}_W + \sum_{l=1}^N \mathcal{H}_R^l
+\sum_{l=1}^N \mathcal{V}_{WR}^l \nn    \\  
{\rm where}~~~\mathcal{H}_W &=& -\sum_{l=1}^{N-1}\g~(c^{\dag}_l c_{l+1}+
 c^{\dag}_{l+1} c_l  ) \nn \\
\mathcal{H}_R^l &=& -\gamma_l~\sum_{\alpha=1}^{\infty}~( c^{l\dag}_{\alpha} 
 c^l_{\alpha+1} +c^{l\dag}_{\alpha +1} c^l_{\alpha})~~~~l=1,2..N \nn \\
\mathcal{V}_{WR}^l &=& - \gamma'_l~ ( c^{l\dag}_{1} c_l + c^{\dag}_l c^l_{1} )~~~~l=1,2...N~. 
\end{eqnarray}
Here $ c_l$ and $ c_\alpha^l $ denote respectively operators on the
 wire  and on the $l^{th}$ reservoir. The Hamiltonian of wire is
 denoted by $\mathcal{H}_W$, that of the $l^{\rm th}$ reservoir
 by $\mathcal{H}^l_R$ and the coupling between the wire and the
 $l^{\rm th}$ site is $\mathcal{V}^l_{WR}$.  The coupling between the
 reservoirs and the wire is controlled by the parameters $\gamma'_l$.

We briefly indicate the steps leading to generalized quantum Langevin
equations of motion for the wire variables. We assume that for $t \leq
t_0$ the reservoirs are  disconnected from the wire. Each reservoir is in
equilibrium at a specified temperature $T_l$ and chemical potential
$\mu_l$. At time $t_0$ we connect all the reservoirs to the wire and
we are interested in the steady state properties of the wire. 
For $t > t_0$, the Heisenberg equations of motion for the wire and reservoirs variables are:
\begin{eqnarray}
\dot{c}_l&=& \f{i \g}{\hbar}(c_{l-1} + c_{l+1}) + \f{i \g_l'}{\hbar} 
c^l_{1} ~~~~{\rm for} ~~ l = 1,2...N~,  \label{eqmW}\\ 
\dot{c}^l_{\alpha}&=& \f{i \g_l}{\hbar} ({c^l}_{\alpha
  -1}+{c^l}_{\alpha +1}) ~~~~{\rm for}~~~ \alpha  = 2,3,...\infty~,~~
l=1,2,...N   \label{eqmR1}\\ 
\dot{c}^l_{1} &=& \f{i \g_l}{\hbar} c^l_{2}+ \f{i \g'_l}{\hbar} c_l~~~
    {\rm for}~~ l = 1,2...N \label{eqmR2}
\end{eqnarray} 
and we have taken $c_0=c_{N+1}=0$.  
The equation of motion of the wire variables Eq.~(\ref{eqmW}) involves
the reservoir variable $c^l_1$ and we will try to eliminate this.
We note that the equation of motion of each of the $N$ reservoirs,
given by Eq.~(\ref{eqmR1},\ref{eqmR2}), is a set of 
linear equations with an inhomogeneous part given by $ i \gamma'_l c_l/\hbar
$ . We can solve these equations of motion using the single
particle Green's function of the reservoirs which is given by  
 $g^l(t)=-i \theta(t) e^{-i H^l t/\hbar} $ where $ H^l$ is the
single-particle Hamiltonian of the  $l^{\rm th}$ reservoir and
$\theta(t)$ the Heaviside step function.  
We finally find that the solution for the boundary site on the $l^{\rm
  th}$ reservoir is given by (for $t > t_0$)
\bea
 c^l_{1}(t)= i \sum_{\alpha = 1}^{\infty}g^{l+}_{1\alpha}(t-t_0)
 c^l_{\alpha}(t_0) - \int_{t_0}^{\infty}~dt'~ g^{l+}_{1,1}(t-t')~\f{\g'_l}{\hbar}~
 c_l(t')  
\end{eqnarray}
Plugging this into the equation of motion Eq.~(\ref{eqmW}) of wire
variables, we get 
\begin{eqnarray}
\dot{c}_l(t)&=& \f{i \g}{\hbar} (c_{l-1}+c_{l+1}) -i\n_l -i \int_{t_0}^\infty
dt' \Sigma^+_l (t-t') c_l (t') \label{gle} \\
{\rm where}~~~\n_l(t) &=&  - \f{i \g_l'}{\hbar}~ \sum_{\alpha =
  1}^{\infty}  g^{l+}_{1\alpha}(t-t_0)~c^l_{\alpha}(t_0) \nn \\
\Sigma^+_l(t) &=& (\f{\g_l'}{\hbar})^2~g^{l+}_{1,1}(t) \nn
\end{eqnarray}
This is in the form of a generalised quantum Langevin equation where we 
identify $\eta_l$ as noise from the $l^{\rm th}$ reservoir and the
last term in Eq.~(\ref{gle}) is the dissipative term. The noise
depends on the reservoir's initial distribution which we have chosen
to be an equilibrium distribution. The properties of the noise is
written most conveniently in the frequency domain. We consider the
limit $t_0 \to - \infty$. Let us define the
Fourier transforms $\tc_l(\om)= (1/2\pi) \int_{-\infty}^\infty dt e^{i
  \om t} c_l(t)$, 
$g^{l+}(\om)=\int_{-\infty}^\infty dt e^{i \om t} g^{l+}(t)$, $\tn_l(\om)=(1/2\pi)~ \int_{-\infty}^\infty dt
e^{i \om t} \n_l(t)$ and $\Se^+_l(\om) = (\g'_l/\hbar)^2
g^{l+}_{1,1}(\om)$. Let us also use the definition $\G_l (\om)=
-Im[\Se^+_l]/\pi= (\g'_l/\hbar)^2 \rho_l(\om)$ where $\rho_l (\om)$ is
the local density of states at the first site ($\alpha=1$) on the
$l^{\rm th}$ reservoir. With these definitions it is easy \cite{AbhishekSen06}  to show that 
the noise-noise correlations are given by 
\begin{eqnarray}
\la \tn_l^\dag (\om) \tn_m(\om') \ra = \G_l(\om)~ f(\omega,~\mu_l,~
T_l)~\delta (\omega -\omega')~\delta_{lm}~,  \label{nncor}
\end{eqnarray}
where $f(\om,\mu,T)=1/\{exp[(\hbar \om-\mu)/k_B T]+1\}$ is the Fermi
distribution function.

Taking Fourier transform of the equation of motion Eq.~(\ref{gle}) we thus get the
following steady state solution
\bea
\tc_l(\om)&=&\sum_{m=1}^N ~G^+_{lm}(\om) ~\tn_m(\om) \label{sol} \\
{\rm where}~~~G^+&=& \f{\hbar}{\g}Z^{-1} \nn \\
{\rm and}~~~Z_{lm}&=&\f{\hbar}{\g}(\omega-\Se^+_l) ~\delta_{lm}+\delta_{l,m-1}+\delta_{l,m+1}~. \nn  
\eea
As shown in [\onlinecite{AbhishekSen06}] $G^+(\om)$ is basically 
the Green's function of the full system (wire and reservoirs) and for
points on the wire can be written in the form $G^+(\om)=
[\om-H_W/\hbar-\bar{\Se}^+]^{-1}$ where $H_W$ is the single particle Hamiltonian
of the wire while $\bar{\Se}^+$, defined by its matrix elements
$\bar{\Se}^+_{lm}=\Se^+_l \delta_{lm}$, is a self-energy correction arising
from the interaction with the reservoirs.  
We will be interested in particle and energy currents in the system. The
corresponding operators are obtained by defining particle and energy
density operators and obtaining their  continuity equations
\cite{AbhishekShastry03}. 
The particle density is defined on sites while the energy density is
defined on bonds.
We will be interested in currents both inside the wire and between the wire
and reservoirs. Let us define $j^p_l$ as the particle current between sites
$l~,l+1$ on the wire and $j^u_l$ as the energy current between the bonds
$(l-1,l)$ and $(l,l+1)$. Also we define $j^p_{w-l}$ as the particle
current from the  wire to the $l^{\rm th}$
reservoir and similarly $j^u_{w-l}$ is the energy current from the
wire to the $l^{\rm th}$ reservoir. These are given by the following
expectation values:
\bea
j^p_l &=&\f{i \g}{\hbar} \la ~c_{l+1}^\dag c_l- c_l^\dag c_{l+1}~ \ra
\nn \\
j^u_l &=& \f{i \g^2}{\hbar} \la ~c_{l-1}^\dag c_{l+1}- c_{l+1}^\dag
c_{l-1}~ \ra \nn \\
j^p_{w-l} &=& \f{-i \g_l'}{\hbar} \la~ c_l^\dag c_1^l - {c_1^l}^{ \dag}
c_l ~\ra \nn \\
j^u_{w-l} &=& \f{i \g \g_l'}{\hbar} \la~ (c_{l+1}^\dag +c_{l-1}^\dag)~
c_1^l  -{c^l_1}^{\dag}   (c_{l+1} +c_{l-1})~ \ra ~.\nn
\eea
Using the general solution in Eq.~(\ref{sol}) and the noise properties
in Eq.~(\ref{nncor}) we can evaluate the above expressions and find
\bea
j^p_l&=&\sum_{m=1}^N \f{-i \g {\g'_m}^2}{\hbar^3} \int_{-\infty}^\infty d \om
(~G^+_{lm} G^-_{m l+1}-G^+_{l+1 m}G^-_{m l}~)~ \rho_m~(f_l-f_m)
\label{jpw} \\
j^u_l&=&\sum_{m=1}^N \f{i \g^2 {\g'_m}^2}{\hbar^3} \int_{-\infty}^\infty d \om
(~G^+_{l-1m} G^-_{m l+1}-G^+_{l+1 m}G^-_{m l-1}~)~ \rho_m~(f_l-f_m)
\label{juw} \\
j^p_{w-l}&=& \sum_{m=1}^N \f{1}{2 \pi}
\int_{-\infty}^\infty d \om \mT_{lm}~(f_l-f_m) \label{jpwl}
\\
j^u_{w-l}&=&  \sum_{m=1}^N \f{1}{2 \pi} 
\int_{-\infty}^\infty d \om ~\hbar \om~\mT_{lm}~(f_l-f_m) \label{juwl}~,
\eea
where $G^-_{lm}={G^+_{ml}}^*$ and $\mT_{lm}=4 \pi^2 {\g'_l}^2 {\g'_m}^2 \rho_l \rho_m |G_{lm}^+|^2/\hbar^4$
can be shown to be the transmission probability of a wave from the
$l^{\rm th}$ to the $m^{\rm th}$ reservoir.

\section{  Self-consistent determination of chemical potential profile }
\label{sec:selfcons}

We consider the case where the wire is held in a fixed temperature
field specified by the temperatures, $T_l,~l=1,2...N$, of the $N$
reservoirs. We will consider a small temperature difference and assume 
that the applied temperature field has the linear form 
\bea
T_l=T_L+\f{l-1}{N-1} \Delta T~, \nn
\eea
where $\Delta T= T_R-T_L$. The chemical potentials at the
ends of the wire are specified by the conditions $\mu_1=\mu_L$ and
$\mu_N=\mu_R$. The $N-2$ side reservoirs $l=2,3...N-1$ are included to simulate other degrees of
freedom present in a real wire and the requirement of zero net particle
current into these reservoirs self-consistently fixes the values of their chemical potentials.   
Thus the chemical potentials $\{ \mu_l \}$ for $l=2,3...N-1$ are obtained by
solving the following set of $N-2$ equations:
\bea
j^p_{w-l}=0 ~~~{\rm for}~~~l=2,3,...N-1~, \label{SCeq}
\eea 
with $j^p_{w-l}$ given by Eq.~(\ref{jpwl}). Once the chemical
potential profile is determined, we can use Eqs.~(\ref{jpw},\ref{juw})
to determine the particle and heat currents in the system while
Eq.~(\ref{juwl}) gives the heat exchange with the environment (side reservoirs).

In general the set of equations Eq.~(\ref{SCeq}) are nonlinear and 
difficult to solve analytically. We will henceforth consider the
low temperature and linear response regime 
where the applied chemical potential difference $\Delta
\mu=\mu_R-\mu_L$ and the temperature difference $\Delta T$ are both small.
More specifically we shall assume $\Delta \mu << \mu_{L,R},~\Delta T
<< T_{L,R}$ and $k_B T_{L,R} << \mu_{L,R}$. For simplicity we restrict
ourselves to the following choice of parameters: $\g_l=\g'_1=\g'_N=\g$
for $l=1,2...N$ and $\g'_l=\g'$ for $l=2,3...N-1$. Thus all the
reservoirs will have the same Green's function and density of states
and we will use the notation $g^{l+}_{1,1}(\om)=g^+(\om)$ and $\rho_l(\om)=\rho(\om)$.

Making Taylor expansions of the Fermi functions $f(\om,\mu_l,T_l)$
about the mean values $\mu=(\mu_L+\mu_R)/2$ and $T=(T_L+T_R)/2$, we
find that, in the linear response regime, Eq.~(\ref{SCeq}) reduces to
the following set of equations:
\bea
j^p_{w-l}=\sum_{m=1}^N \f{1}{2 \pi \hbar}~[~\mT_{lm}~(\mu_l-\mu_m)+
  \f{\pi^2 k_B^2 T }{3\hbar}~\mT'_{lm} (T_l-T_m)~]=0~~~{\rm
  for}~~~l=2,3...N-1~, \label{jpwlLR}
\eea 
where $\mT_{lm}$ and $\mT'_{lm}=d \mT_{lm}/d\om$ are evaluated  at $\om=\mu/\hbar$.
These are linear equations in $\{ \mu_l \}$ and are straightforward to
solve numerically. We can then use Eq.~(\ref{jpw}) and Eq.~(\ref{juw})
to find the particle and heat current. The local heat current in the
wire is given by $j^q_l=j^u_l-\mu_l j^p_l$. In the linear response
regime we find
\bea
j^p_l &=& \f{-1}{2 \pi \hbar} \sum_{m=1}^N [\mF_{lm}~ (\mu_l-\mu_m) +
  \f{\pi^2 k_B^2 T}{3 \hbar} \mF'_{lm}~ (T_l-T_m)] \nn \\
j^q_l &=& \f{-1}{2 \pi \hbar}  \sum_{m=1}^N [~  \f{\pi^2 k_B^2 T^2}{3 \hbar}
  \mF'_{lm}~ (\mu_l-\mu_m) +\f{\pi^2 k_B^2 T}{3}\mF_{lm} ~(T_l-T_m)~]~, \label{jwLR}
\eea
 where $\mF_{lm}= (2 \pi i\g \g'^2/\hbar^3) ~(~G^+_{lm} G^-_{m l+1}-G^+_{l+1 m}
 G^-_{ml}~)~\rho$ and $\mF_{lm}, \mF'_{lm}$ are evaluated at $\om=\mu/\hbar$.
The heat loss from the wire to the reservoir can be obtained
using Eq.~(\ref{juwl}). As we shall see later this heat loss is a
second order effect and therefore we will keep terms up to second order
in the expansion. We then get
\bea
&& j^q_{w-l}=\f{1}{2 \pi \hbar} \sum_{m=1}^N \left[-\f{\pi^2 k_B^2 T^2}{3 \hbar}
  \mT'_{lm}~ (\mu_l-\mu_m) -\f{\pi^2 k_B^2 T}{3}\mT_{lm} ~(T_l-T_m)~ \right.
\nn \\
 &+& \left. \f{1}{2}\mT_{lm}~(\mu_l-\mu_m)^2+\f{2\pi^2 k_B^2 T}{3\hbar}~ \mT'_{lm}
~(\mu_l-\mu_m)(T_l-T_m)+ \f{\pi^2 k_B^2}{3}~ \mT_{lm}~ (T_l-T_m)^2~\right].  \label{jqwlLR}
\eea
In the next section we will the consider the case of a long wire ($N
\to \infty$) and
consider particle and heat transport in the presence of applied
chemical potential and temperature gradients. Later, for an
isothermal system, we will consider  finite systems and discuss the
transition from coherent to Ohmic transport.

\section{Long wire with applied chemical potential and temperature
  gradients}
\label{sec:solu}

Let us  first evaluate the matrix elements $\mT_{lm}(\om)$. This
involves $\rho(\om)$ and $G^+_{lm}(\om)$. As discussed before $\rho(\om)$ is
the local density of states at the boundary site of a semi-infinite
one-dimensional chain and is given by 
$\rho(\omega)=({\hbar}/{(\pi \g)}[ 1-{\hbar^2 \omega^2}/{(4 \g^2)}]^{1/2}$ for
$|\hbar \om |< 2 \gamma$ 
and zero elsewhere. For lattice points in the bulk of the wire, \emph{i.e} points which
are at a distance $>> \ell= 1/\alpha_R$ from the boundaries of the wire we
find (see Appendix~\ref{appG}) $ G^+_{lm}= {(-1)^{l+m}\hbar~
  e^{-|l-m|\al}}/(2 \g \sinh{\al})$. 
We now try the following solution for the self-consistent equations
given by Eq.~(\ref{jpwlLR}):
\bea
\mu_l=\mu_L+\f{l-1}{N-1}\Delta \mu~. \label{musol}
\eea
Using
the fact that 
$\sum_{m=-\infty}^{m=\infty} (l-m)~ e^{-|l-m|\al}=0$, we see that
the self-consistent equations are satisfied
for all points $l$ in the bulk of the wire (up to corrections
which become exponentially small with the distance from the
boundaries). Close to the boundaries the chemical potential variation
is no longer linear. Here we focus on the limit where $N$ is very
large and the linear solution in Eq.~(\ref{musol}) is accurate in the
bulk of the wire. 

We will now use this solution to evaluate the
various currents in the wire given by Eq.~(\ref{jwLR}) and the heat
loss from Eq.~(\ref{jqwlLR}). We evaluate these
currents at points $l$ in the bulk of the wire and (since $G^+_{lm}$
decays exponentially with distance) do not need the correct form of
$\mu_l$ at the boundaries. We also find, as expected,  that the
currents are independent of $l$. They have the expected linear
response forms:
\bea
j^p&=& -L_{11} \nabla \mu -L_{12} \nabla T \nn \\
j^q&=&-L_{21} \nabla \mu -L_{22} \nabla T~, \nn
\eea
where $\nabla \mu=\Delta \mu/N$, $\nabla T=\Delta T/N$ and the
various transport coeficients are given by
\bea
L_{11}&=&\f{1}{2 \pi \hbar} \sum_{m=-\infty}^\infty
\mF_{lm}(\mu/\hbar)~ (l-m) = \f{1}{\pi \hbar}~\f{\sin^2{\al_I}  \coth{\al_R} }
{\cosh2{\al_R}-\cos2{\al_I}}~,\label{l11} \\
L_{12}&=&\f{\pi^2 k_B^2 T}{3}~ \f{dL_{11}}{d \mu}~, \label{mott} \\ 
L_{21}&=&T~L_{12}~,\label{onsag}\\
L_{22}&=&\frac{{\pi}^2 k_B^2 T}{3}~L_{11}~,\label{weidfr} 
\end{eqnarray}
where $\al_R$ and $\al_I$ are respectively the real and imaginary
parts of $\al$ and are all calculated at $\om=\mu/\hbar$.  In deriving
the above form of $L_{11}$ we have used the relation $\rho=2 \g \hbar
\sinh{\al_R} \sin{\al_I}/(\pi {\g'}^2)$. In the parameter regime we
are looking at, it follows that both Ohm's law and Fourier's law are
valid, with  the electrical and thermal
conductivities given by $\sigma=e^2 L_{11}$ and
$\kappa=L_{22}$. 
We note that 
Eq.~(\ref{onsag}) gives the Onsager reciprocity relation.
This is usually derived within linear response theory and follows from
time reversal 
invariance of the microscopic equations of motion. We
also find from Eq.~(\ref{weidfr}) that the
Weidemann-Franz relation is satisfied. This
relation states that the ratio of the thermal conductivity and the electrical
conductivity 
is linearly proportional to the temperature with a 
universal constant of proportionality given by ${\pi^2 k_B^2
}/{(3e^2)}$. For metals a derivation of this relation using
semiclassical tramsport theory and within the relaxation time
approximation can be found in  [\onlinecite{ashcroft}]. The validity of this
relation requires that inelastic processes can be neglected ( see
discussion in [\onlinecite{ashcroft}] ). However we find that the
relation continues to be valid in our model even though scattering is
inelastic (since there is energy dissipation into the side reservoirs).  

\begin{figure}[t]
\begin{center}
\includegraphics[width=12.0cm]{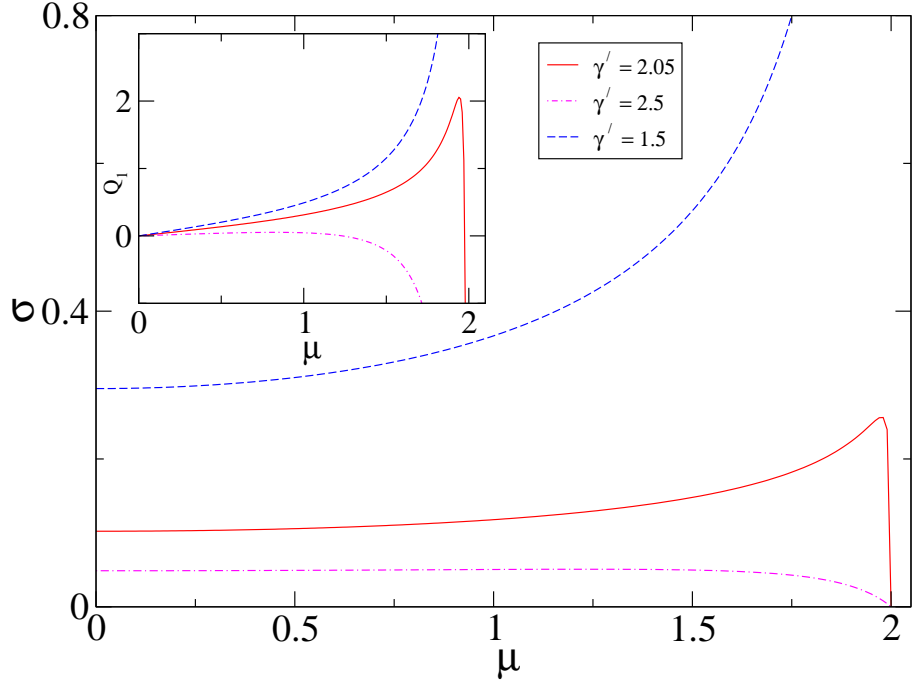}
\end{center}
\caption{ Plot of the conductivity and thermopower $Q_1=
3eQ/(\pi^2 k_B^2 T)$ (Inset) as 
  functions of the Fermi level $\mu$ for different values of the
  inelasticity parameter $\gamma'$. They are plotted in 
  units of $e^2/\pi \hbar$ and $\gamma$ respectively.
}
\label{condvsmu}
\end{figure}

From Eq~(\ref{mott}) we find that  the  Mott formula for the thermopower
holds \cite{mott,lunde05}. This is given by
\bea
Q=\f{L_{12}}{e L_{11}}=\f{\pi^2 k_B^2 T}{3 e} \f{1}{\sigma}\f{d
  \sigma}{d \mu}~.
\eea
Recently [\onlinecite{lunde06}] have reported an
interesting resonance, arising due to electron-electron  interactions,
observed in the thermopower as a function of the Fermi energy. 
We investigate if there are any interesting features in the
dependence of $Q$ on $\mu$ in our model. In Fig~(\ref{condvsmu}) we
plot the conductivity and the thermopower $
3eQ/(\pi^2 k_B^2 T)=d(\ln \sigma)/d\mu$ as a function of $\mu$ for
different values of the coupling constant $\gamma'$. Surprisingly we
find that for a range of  values of the inelasticity parameter
$\gamma'$ there is a peak in the thermopower as a function of the
Fermi energy. 

Let us now look at the heat exchanges given by
Eq.~(\ref{jqwlLR}). In the long-wire limit, the condition of zero
particle currents into the side reservoirs, Eq.~(\ref{jpwlLR}),
implies that $\sum_m \mT_{lm} (l-m)=\sum_m \mT' (l-m)=0$.  Hence the
terms linear in $\nabla \mu$ 
and $\nabla T$  in Eq.~(\ref{jqwlLR}) vanish, and only the second
order terms contribute significantly. Let 
us first consider  the coefficient of the term containing $(\nabla
\mu)^2$ which is given by
\bea
\f{1}{4\pi\hbar} \sum \mT_{lm} ~(l-m)^2~.
\eea
Evaluating the sum we find that it is exactly equal to
$L_{11}$. Determining the other terms in Eq.~(\ref{jqwlLR}) we find
that the net heat loss per unit length (or from every bulk site) of the wire  is given by:
\bea
j^q=L_{11}~ (\nabla \mu)^2 + \f{4 \pi^2 k_B^2 T}{3}~ \f{d L_{11}}{d \mu}~ (\nabla
\mu) (\nabla T) + \f{2 \pi^2 k_B^2}{3} ~L_{11} (\nabla T)^2~.
\eea
The first term corresponds to the expected Joule heat loss in a wire
and is always positive. The second term can be of either sign and can
be identified to be the Thomson effect which corresponds to heat
exchange that occurs in a wire (in addition to the Joule heat) when an
electric current flows across a temperature gradient. 

Finally we check for local thermal equilibrium in the wire. A
requirement of local equilibrium would be that the local density $n_l$
at the point $l$ in the nonequilibrium state should be the same as the 
density $n^{eq}_l$ at the point if the entire wire was kept in
equilibrium at a chemical potential $\mu_l$ and temperature $T_l$. It is easy to evaluate
$n_l$ and $n^{eq}_l$ and we find:
\bea
n_l-n^{eq}_l= \sum_{m=1}^{N} \f{{\g'}^2}{\hbar^2} \int_{-\infty}^\infty d \om
|G^+_{lm}(\om)|^2 ~\rho(\om)~[~f(\om,\mu_m,T_m)- f(\om,\mu_l,T_l)~]~, \nn
\eea
which in the linear response regime, gives
\bea
n_l-n^{eq}_l= \f{{\g'}^2}{\hbar^3} ~ \sum_{m=1}^{N}
\left[~|G^+_{lm}(\mu/\hbar)|^2 ~\rho(\mu/\hbar)~ (\mu_l-\mu_m) + \f{\pi^2
    k_B^2 T }{3}~ \f{d}{d\mu}[
    |G^+_{lm}(\mu/\hbar)|^2 \rho(\mu/\hbar)]~(T_l-T_m)~\right]~.\nn
\eea
For our linear profiles of temperature and chemical potential and the
form of $G_{lm}$  it is clear that, for all bulk points, the above
difference vanishes (upto order $O(1/N)$). Thus we see that the local
densities are consistent with the assumption of local equilibrium.

\section{Current flow in wire in isothermal conditions: finite size effects } 
\label{sec:finite}

In this section we look at finite length wires. We first solve
Eq.~(\ref{jpwlLR}) numerically to determine the chemical potential
profile and then estimate the current in the wire
using Eq.~(\ref{jwLR}). We also look at the local  heat dissipation at
all points on the wire. 

\begin{figure}[t]
\begin{center}
\includegraphics[width=12.0cm]{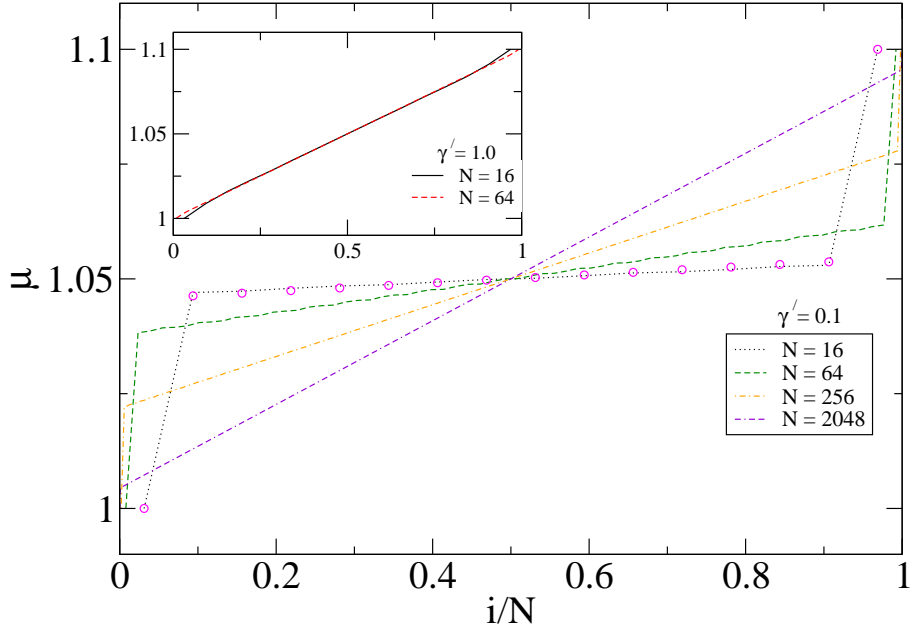}
\end{center}
\caption{ Plot of the chemical potential profile $\mu_i$ as a function
  of the scaled length $i/N$ for different values of $N$ and with
  $\gamma'=0.1$. The points denoted by circles correspond to the
  approximate solution given in Eq.~(\ref{muscal}). The inset shows
  the chemical potential profile for   $\g'=1.0$. 
}
\label{muprof}
\end{figure}

In our numerical calculations we have chosen the parameter
values $\mu_L=1.0, \mu_R=1.1, \gamma=1.0$ and have considered
different values of the dissipation strength $\g'$ and different
system sizes $N$. 
In Fig.~(\ref{muprof}) we plot the chemical potential profile,  for
different system sizes, for a small value of the dissipation
($\gamma'=0.1$). We see that  as we go to larger system sizes,
chemical potential profile changes, from a flat profile
with large jumps at the boundaries, to a smooth linear profile. For a
larger dissipation parameter ($\g'=1.0$) we see [inset of
Fig.~(\ref{muprof})] that a smooth linear profile is obtained even for small
system sizes. 
The limit of weak dissipation was studied in [\onlinecite{Pastawski90}]
. Following them we find that for  $\g'/\g <<1$ a very
good approximation for the transmission coefficients $\mT_{lm}$, for any system size,  is
given by
\bea
\mT^+_{lm}=\f{\pi^2 {\g'_l}^2 {\g'_m}^2 \rho^2}{{\hbar}^2 \g^2}
e^{-\f{2 |l-m|}{\ell}}~. \label{tlmapp}  
\eea
where $\ell=1/\alpha_R \approx 2\g^2/\g'^2$.
Note that, for $l=2,3,...N-1$, $\mT_{1l}$ and $\mT_{lN}$ are
$O({\g'}^2)$ while $\mT_{lm}$ for $m=2,3,...N-1$ are
$O({\g'}^4)$. We then find that, for $\ell >> 1$ and $N >> 1$, the
following chemical potential profile provides a good approximate
solution of the self-consistent equations:
\bea
\mu_1&=&\mu_L,~~\mu_N=\mu_R \nn \\
\mu_l&=& \mu_L-\delta-\f{2\delta}{\ell}(l-2)~~~
    {\rm for}~~ l = 2,3,...N-1\label{muscal} \\
{\rm where}~~~\delta&=& \f{\mu_L-\mu_R}{2(1+N/\ell)} \nn
\eea
Plugging in this  solution into the
self-consistency equations $\sum_m \mT_{lm} (\mu_l-\mu_m)=0$ with
$\mT_{lm}$ given by Eq.~(\ref{tlmapp}) we can explicitly verify that
these are satisfied upto corrections of order $1/\ell$.    
In Fig.~(\ref{muprof}) we have  plotted the above  solution for
 system size $N = 16$ and find an excellent agreement with
the numerical result (for larger system sizes the fit agreement
becomes better).  

The above solution leads to the following result for the current:
\bea
\f{j N}{\Delta \mu L_{11}}=\f{1}{1+\ell/N}~, \label{scal}
 \eea
where $L_{11}$ is the ohmic conductivity of the wire given by
Eq.~(\ref{l11}) and  we have normalized the current such that the $N\to \infty$ limit
gives a constant value independent of $\g'$.

\begin{figure}[t]
\begin{center}
\includegraphics[width=12.0cm]{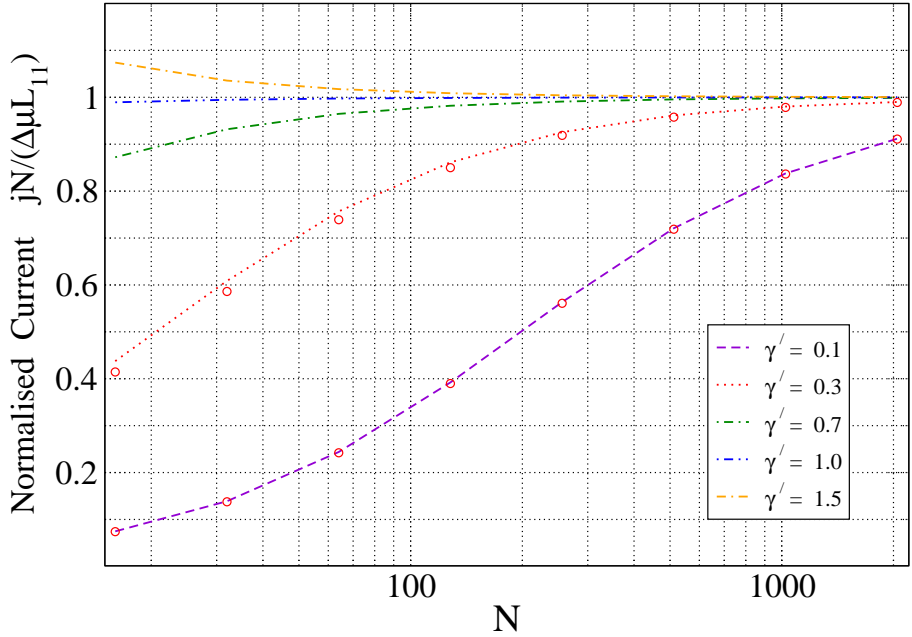}
\end{center}
\caption{Plot of the normalized current versus system size for
  different values of the dissipation constant $\g'$. The  points
  denoted by circles 
  correspond to the analytic scaling form given in Eq.~(\ref{scal}). 
}
\label{jvsl}
\end{figure}

We have also looked at  the transition from coherent to Ohmic transport for general
values of the dissipation parameter $\g'$. 
In Fig.~(\ref{jvsl}) we plot the scaled current $j N/ (\Delta \mu L_{11})$ as a
function of system size. We find that in general for any $\g'/\g < 1$ the data can be
fitted quite accurately to the form in Eq.~(\ref{scal}) 
with $\ell=1/\al_R$ which can be interpreted as a coherent length scale. 
For $\g'/\g > 1 $ we find that there is no coherent regime and the
approach to the asymptotic limit has a different form.

{\emph{Persistent random walk model}}: It is possible to understand the
various aspects of the intermediate regime within a simple classical
Drude-like  framework of right moving and left moving electrons moving
in fixed  directions but with a small probability of inter-conversion. 
We consider the case where the left reservoir is kept at a
chemical potential $\mu+\Delta \mu$ and the right reservoir is at
$\mu$. At the low temperatures being considered electron transport is
basically due to the electrons close to 
the Fermi level and we can focus on the electrons within the energy
gap $\Delta \mu$ in the left reservoir. Let the density of these
electrons inside the left reservoir be $2\rho_L$ and this consists of
an equal proportion of right moving electrons with velocity $v_F$ and
left movers with velocity $-v_F$. In the right reservoir the density
of both left and right movers in the energy window $\Delta \mu$ is
zero. Inside the wire the presence of the side-reservoirs allows a
right mover to 
be converted to a left mover with some probability. 
We now present the following random walk model to incorporate the
above basic idea. The model consists of a lattice of $N$ sites with a
density $\rho_l^+$ of right movers and $\rho_l^-$ of left
movers at all sites $l=1,2...N$. We impose the boundary
conditions $\rho_1^+=\rho_1^-=\rho_L$ and $\rho_N^+=\rho_N^-=0$. At
sites $l=2,3...N-1$  the particles move according to the following
rules: with probability $p$ a right mover at site $l$ moves to $l+1$
and, with probability $1-p$ it transforms to a left mover and moves to
site $l-1$. Similarly, with probability $p$ a left mover at site $l$
moves to $l-1$ and, with probability $1-p$ it transforms to a right mover
and moves to $l+1$.  
At sites $l=1,N$, the right mover always moves to the right
and the left mover moves to the left. It is then straightforward to
write discrete time-evolution equations for the density fields
$\rho_i^+(t)$ and $\rho_i^-(t)$. Choosing a lattice length scale $a$
and a microscopic time scale $\tau$ we obtain, in the continuum limit 
\bea
\f{\p \rho^+(x,t)}{\p t}&=&-v \f{\p \rho^+}{\p x} -\alpha
(\rho^+-\rho^-) \nn \\     
\f{\p \rho^-(x,t)}{\p t}&=&v \f{\p \rho^-}{\p x} +\alpha
(\rho^+-\rho^-)~. \label{diffeqn}
\eea
where $v=a p/\tau$ can be identified with the Fermi velocity $v_F$ and
$\alpha=(1-p)/\tau$ gives the scattering rate (Note that the continuum
limit requires taking $a\to 0$, $\tau \to 0$ and $p\to 1$ keeping $v$
and $\alpha$ finite). We obtain a length scale $v/\alpha$ which we
tentatively identify with the scattering length $\ell$ introduced
earlier. The boundary conditions for the above equations are
$\rho^+(x=0)=\rho_L$ amd $\rho^-(x=L)=0$, where $L=Na$. These give the
following steady state solution for Eq.~(\ref{diffeqn}):
\bea
\rho(x)&=&\rho^+(x)+\rho^-(x)=2\rho_L-\delta'-\f{2 \delta'}{\ell} x  \\
{\rm where}~~~\delta' &=& \f{2 \rho_L}{2(1+L/\ell)}
\eea  
is the density jump at the boundaries. This immediately leads to
Eq.~(\ref{muscal}) once we note that $\rho(x)-2\rho_L \propto
\mu_l-\mu_L$. The current in the wire is given by
\bea
J=v [\rho^+(x)-\rho^-(x)]=\f{\ell v \rho_L}{L(1+\ell/L)}
\eea  
which again leads to the result in Eq.~(\ref{scal}) after we make the
appropriate identifications.

An interesting question that is often asked in the context of
mesoscopic transport is: {\emph{where is the dissipation }}
\cite{Datta05} ? In the
case of Ohmic transport, dissipation, through Joule heat loss, takes
place in the bulk of the wire. On the other hand for coherent transport
there is no dissipation in the bulk of the sample and the only
dissipation is at the contacts (or into the leads). 
This difference between Ohmic and coherent transport can be 
demonstrated in our model by an explicit calculation of the local heat loss
at all points on the wire. Using Eq.~(\ref{jqwlLR}) we calculate the
fraction of the total heat loss that occurs at the contacts 
$j^q_C=j^q_{w-1}+j^q_{w-N}$ and the bulk heat loss given by
$j^q_B=\sum_{l=2}^{N-1} j^q_{w-l}$. Note that the total dissipation is
given by $\sum_{l=1}^N j^q_{w-l} = j^p \Delta \mu $ which easily
follows from using the condition $\sum_{l=1}^N j^u_{w-l}=0$. The
following table shows the contact and bulk heat losses for different
system sizes and with $\g'=0.1$. In this case $\ell \approx 200$. We
see clearly that for $N << \ell$ dissipation occurs mostly in the
contacts to the 
leads while for $N >> \ell$ dissipation occurs in the bulk of the wire. 
Note that the heat is eventually dissipated into the reservoirs and is
possible even in a steady-state scenario because of the infinite size
of the reservoirs.  
\begin{table}
\begin{tabular}{| r | c | c | }
\hline
\hline
$L$  &  $j^q_C$ & $j^q_B $ \\
\hline
  16 & $0.9652$ & $0.0348$           \\
  64 & $0.8518 $ & $0.1482$    \\
  256 & $0.5425 $ & $0.4575 $    \\
  512 & $0.3522 $ & $0.6478 $    \\
  1024 & $0.2049 $ & $0.7951 $    \\
 2048 & $0.1115 $ & $0.8885 $         \\
\hline
\hline
\end{tabular}
\end{table}

\section{Discussion}
\label{sec:disc}

An interesting aspect of the present study arises if we compare it
with studies of heat transport by phonons in oscillator chains. A big
question there has been to find the necessary conditions on a model of
interacting particles required for the validity of Fourier's law of
heat conduction \cite{bonetto00}. As a result of a large number of
studies it now appears 
that heat conduction in one dimensions is anomalous and Fourier's law
is not valid for  momentum conserving models \cite{lepri03}. However there are
stochastic models where one can exactly demonstrate the validity of Fourier's
law. In one such model inelastic scattering of phonons take place by
an exact analogue of the B\"uttiker probes. In this model, first
proposed in [\onlinecite{Bolsterli70}], and solved exactly recently in
[\onlinecite{Bonetto04,AbhiDib06}] 
, each site on a harmonic lattice is connected
to a heat reservoir whose temperature is fixed self-consistently by
the condition of zero heat current. Just as Fourier's law can be shown
to hold in this model, here we have shown that both Fourier's law and
Ohm's law are valid in the present tight-binding model. We have also
been able to explicitly demonstrate local thermal equilibrium and
various other linear response results. One other model where such a
demonstration has been made in a clear way is the work by Larralde et
al \cite{Larralde03} on the Lorentz gas model. One other point to note is, as shown in [\onlinecite{AbhishekSen06}]
, the treatments of electron and phonon transport can be
done in  a very similar way using the formalism of quantum Langevin
equations and nonequilibrium Green's function.

In this paper we have extended the calculation of D'Amato and
Pastawski by studying the finite temperature case and considering transport
of both particles and heat in a tight-binding chain. We have studied
both the Ohmic and ballistic regimes. It has been shown that a simple
Drude-like model of persistent random walkers can explain many of the
observed features  in the intermediate regime. In the Ohmic regime we
have calculated verious thermoelectric coefficients and find that for
certian values of the inelasticity parameter, the thermopower plotted
as a function of the Fermi energy shows a peak. Finally we have
explicitly computed heat dissipation in the wire. 

While we have only
considered the linear response regime in this paper, the formalism
described here can be used to study the nonlinear regime too.  Also
it can be easily used to study inelastic scattering effects in the
tight-binding model in any dimensions and the reservoirs themselves
can be in any dimensions. Numerical implementations to
study systems with  Anderson type of disorder and systems with
externally applied  magnetic fields can also be done
readily with our approach.  Finally, as pointed out in the
introduction, our  model of inelastic   scattering  also serves as 
a model for voltage probes. An important point in experiments
involving  four terminal resistance measurements on 
 quantum wires, as in  [\onlinecite{dePicciotto01}] for example, is that the
 voltage probe  should be non-invasive. In our model  the coupling to
 the probes  
 can be tuned  and thus can be used to obtain a better understanding
 of the role of probes in such experiments. Also more detailed models
of the probes are easy to incorporate in our approach.  
The quantum Langevin method can be easily used for other models
 of the scattering reservoirs other than the present model where each 
reservoir is a one-dimensional wire. This would basically involve a
change in the form of the self-energy correction.  
An interesting problem is an extension of the present formulation to
include electron-phonon and electron-electron interactions.

\section{acknowledgments}
We thank Diptiman Sen and N. Kumar for useful discussions.

\appendix

\section{Evaluation of Green's Function}
\label{appG}
 To find the Green's function we use the relation
$G^+_{lm}=(\hbar/\gamma)Z^{-1}_{lm}$ where $Z$ is a tridiagonal matrix
with off-diagonal terms all equal to one. The diagonal terms are given
by:
\bea
Z_{11}=Z^+_{NN}&=&A(\om)= \f{\hbar}{\g}~[\om-\f{\g^2}{\hbar^2}
  g^+(\om)] \nn \\
Z_{ll}&=&B(\om)= \f{\hbar}{\g}~[\om-\f{{\g'}^2}{\hbar^2}
  g^+(\om)]  \label{diagelm}~~~{\rm for}~~l=2,3...N-1~.
\eea 
The function $g^+(\om)$ can be obtained from the green function of an
isolated semi-infinite one-dimensional chain and, in the region of
interest here ($|\hbar \om |< 2 \gamma $) is given by
\bea
g^+(\om)=\f{\hbar}{\gamma}\left[ \f{\hbar \om}{2 \g}-i \left(
  1-\frac{\hbar^2 \omega^2}{4 \g^2}\right)^{1/2} \right]~.
\eea
Using standard matrix manipulations we can evaluate the inverse of $Z$ and find
\bea
Z^{-1}_{lm}&=&(-1)^{l+m} \f{D_{l-1} D_{N-m}}{\Delta_N} ~~~~{\rm for}~~m > l \nn \\ 
&=&(-1)^{l+m}\f{D_{m-1} D_{N-l}}{\Delta_N} ~~~~{\rm for}~~m \leq l \label{Zinv} \\ 
{\rm where}~~D_l&=& A Y_{l-1} -Y_{l-2}\nn \\ 
\Delta_N&=&Det[Z]={A}^2  Y_{N-2}-2 A Y_{N-3}+Y_{N-4} \nn \\
Y_l&=&\frac{\sinh[(l+1)\alpha]}{\sinh (\alpha)}  \nn \\
{\rm with}~~ e^{\pm \alpha} &=&\frac{B}{2} \pm (\f{B^2}{4} -1)^{1/2}~.  \nn
\end{eqnarray} 
We will assume that the root $\alpha$ has been chosen such that
$\alpha_R=Re[\alpha] > 0$. Using the above results for the inverse of the
matrix $Z$ we find that for large $N$ the Green's function in the wire
is given by:
\bea
G^+_{lm}= \f{(-1)^{l+m}\hbar}{2 \g \sinh{\al}}~\left[ e^{-|l-m|\al}-\f{(A-e^{\al})}{(A-e^{-\al})} \left(e^{-(l+m-2)\al}+e^{-(2N-l-m)\al}\right)\right]
\eea 


\end{document}